\documentclass[reprint,
tightenlines,
superscriptaddress,
 amsmath,amssymb,
 aps,
pra
]{revtex4-2}

\usepackage{graphicx}
\usepackage{dcolumn}
\usepackage{bm}
\usepackage{braket}
\usepackage{amsmath}
\usepackage{hyperref}
\hypersetup{colorlinks=true,citecolor=blue}

\begin{document}

\title{Non-adiabatic interaction effects in the spectra of ultralong-range Rydberg molecules}

\author{Rohan Srikumar}
\email{rsrikuma@physnet.uni-hamburg.de}
\affiliation{Zentrum für Optische Quantentechnologien, Fachbereich Physik, Universität Hamburg, Luruper Chaussee 149, 22761 Hamburg, Germany}

\author{Frederic Hummel}
\email{hummel@pks.mpg.de}
\affiliation{Max-Planck-Institut für Physik komplexer Systeme, Nöthnitzer Str.~38, 01187 Dresden, Germany}

\author{Peter Schmelcher}
 \email{pschmelc@physnet.uni-hamburg.de}
\affiliation{Zentrum für Optische Quantentechnologien, Fachbereich Physik, Universität Hamburg, Luruper Chaussee 149, 22761 Hamburg, Germany}
\affiliation{The Hamburg Centre for Ultrafast Imaging, Universität Hamburg, Luruper Chaussee 149, 22761 Hamburg, Germany}

\date{\today}

\begin{abstract}
 Ultralong-range Rydberg molecules (ULRM) are highly imbalanced bound systems formed via the low-energy scattering of a Rydberg electron with a ground-state atom. We investigate for $^{23}$Na the $d$-state and the energetically close-by trilobite state, exhibiting avoided crossings that lead to the breakdown of the adiabatic Born-Oppenheimer (BO) approximation. We develop a coupled-channel approach to explore the non-adiabatic interaction effects between these electronic states. The resulting spectrum exhibits stark differences in comparison to the BO spectra, such as the existence of above-threshold resonant states without any adiabatic counterparts, and a significant rearrangement of the spectral structure as well as the localization of the eigenstates. Our study motivates the use of $^{23}$Na ULRM, as a probe to explore vibronic interaction effects on exaggerated time and length scales.
\end{abstract}

\maketitle

\section{Introduction} \label{Sec1}

Rydberg atoms are an important player in modern quantum physics due to their unique and extreme properties. Their size and dipole moment scale as $n^2$, and lifetimes and polarizability scale as $n^3$ and $n^7$, respectively, where $n$ is the principal quantum number \cite{gallagher_1994,sibalic_2018}. They offer a state-dependent interaction strength and enhanced sensitivity to electromagnetic fields, rendering them promising platforms for quantum computing \cite{Saffman_2010,Saffman_2016,Browaeys_2020,Bluvstein_2022,Graham_2022}, external field sensing \cite{Sedlacek2012, Fan_2015,Gordon_2014}, detection of polar molecules \cite{Zeppenfeld_2017,Patsch2022,Zou_2022} and Rydberg-quantum optics \cite{Firstenberg_2016,Peyronel2012,Firstenberg2013,Mandoki_2017}. In 2000, a pioneering article \cite{Greene2000} predicted the formation of ultralong-range Rydberg molecules (ULRM), as a result of attractive scattering interaction of the Rydberg electron with a ground-state perturbing atom. These exotic bound states feature permanent dipole moments of the order of kilodebye, and bond-lengths of the order of micrometers, properties which were unheard of in conventional diatomic molecules. However, recent works have also studied other exotic bound molecular systems with exaggerated properties, formed due to different binding mechanisms like the Rydberg macrodimer \cite{Boisseau_2002,hollerith_2019,Overstreet2009}, and the Rydberg atom-ion molecule \cite{Duspayev_2021,Deiss_2021,Bosworth2022,Raithel_2022,Zuber2022,anasuri_2023}.

This work is dedicated to the diatomic ULRM, bound as a consequence of low-energy electron-atom $s$-wave scattering. The scattering mechanism enables the existence of two distinct classes of molecular states \cite{Greene2000}. The first type are formed when a Rydberg electron of low angular momentum ($l\leq2$) interacts with the neutral perturber. Molecules of the low-$l$ class, exhibit shallow potential wells ($\sim$10~MHz) and a very small net electronic dipole moment \cite{Sadeghpour_2011}. The second type are the trilobite class of molecules \cite{Booth_2015,Bellos_2013}, which arise as the high-$l$ hydrogenic manifold is mixed by the ground state atom \cite{Greene_1987}. The trilobite molecules exhibit much deeper potential wells ($\sim$GHz) and a large permanent dipole moment, and is responsible for garnering much interest to the field of ULRMs. In addition, the use of higher order terms in our scattering interaction has also demonstrated the existence of the butterfly state \cite{Hamilton_2002,Niederprum2016}, which arise as a consequence of $p$-wave shape resonances. Starting from it's experimental observation in 2009 \cite{Bendkowsky2009}, significant progress have been made in the study of ULRM over the past two decades. This includes it's utility, as a probe for spatial correlations in ultracold atomic gases \cite{Whalen_2019}, in precision spectroscopy of negative-ion resonance \cite{Engel_2019}, and in the study of Rydberg impurities in ultracold atomic gases \cite{Schmidt_2016,Sous_2020}. Furthermore, the fine and hyper-fine structure of these molecules \cite{Eiles_2017,Hummel2020,Anderson2014} as well as their behavior in external electric and magnetic fields have been studied \cite{Lesanovsky_2006,Kurz_2014,Hummel_2019,Keiler_2021,Kurz_2013}, and major strides have been made in the experimental exploration of these molecules (see \cite{Shaffer2018,killian2015,killian2020,Pfau2018,Ott2016,Deiglmayer2021,Raithel2014,Engel_2019,Kleinbach_2017,Schalgmuller2016,Raithel_2016,Niederprum2016,Bendkowsky2009,Camargo_2016,Butscher_2011} for examples of experimental investigations on ULRMs).

One of the important theoretical tools used in the analysis of a ULRM is the well-established Born-Oppenheimer (BO) approximation \cite{Born_1927}. As a cornerstone of molecular physics, the BO approximation advocates for the separation of slow nuclear and fast electronic motion allowing us to independently solve the electronic problem in varying geometrical arrangements of the nuclei, to obtain the so-called adiabatic potential energy curves (PECs). However, there is typically an ubiquity of avoided crossings featured in the PECs of the ULRM, and resultingly, non-adiabatic interactions potentially leading to the breakdown of the BO approximation are expected. In the special case of a conical intersection (CI), when the molecular geometry facilitates the degeneracy of two PEC, the non-adiabatic couplings become singular and the adiabatic approximation breaks down completely \cite{koppel_1984,Agostini_2019,Baer_2006}. Recent studies have demonstrated the existence of such CIs in the Rb ULRM, for very specific conditions, when the electron-perturber scattering phase shift divided by $\pi$ is similar in size to the quantum defect i.e.~$\mu_d  \pi \approx \delta_s$ \cite{Hummel2021}. In the context of traditional molecular physics, CIs play an important role in molecular dynamics as they can cause fast non-radiative transitions between electronic states \cite{Arnold_2018,Mabrouk_2020,Martinez2010,Domcke1987}, Jahn-Teller distortions \cite{jahn_1937,Jahn_1958}, and surface hoppings \cite{Herman_1984,Barbatti_2011} to name a few examples. The study of non-adiabatic effects in the vicinity of conical intersections has been proven to be necessary for understanding a wide-range of natural phenomena such as photo stability of DNA \cite{Mario2010,Kang2002}, photoisomerization \cite{Levine2007,Polli2010} and reaction mechanisms involving photosynthesis \cite{Leif_2008}. The ULRM on the other hand, provides us with a platform to explore molecular dynamics in the unique and enormous time-scales of microseconds and on length-scales of micrometers. Hence, the study of non-adiabatic interaction effects, and their contribution to the spectral characteristics in a molecule which features such exaggerated properties, is a challenging and promising direction of research on ULRMs.

In this work, we study the non-adiabatic interaction between two electronic states in $^{23}$Na ULRM, due to the vibronic coupling between them. The electronic level structure of $^{23}$Na leads to avoided crossings and state mixing between the trilobite and the $d$-state PECs (see \cite{Hummel_2022}, for a recent study on vibronic couplings between trilobite and butterfly states). We highlight results for specific $n$-values, with near-degenerate avoided crossings that cause singular vibronic couplings, explained using the concept of CIs in synthetic dimensions. A coupled-channel approach is employed to obtain the vibronic spectra, thereby using the non-adiabatic couplings between the trilobite and $d$-state PECs. We observe features of vibronic interactions in the underlying spectra including scattering resonance states with no adiabatic counterparts. We also use the single-channel Born-Huang approximation, including the non-adiabatic diagonal corrections without the off-diagonal couplings as a comparative resource. The analysis of vibronic and vibrational spectra is used to justify the necessity of a coupled-channel approach to study the $^{23}$Na ULRM.

This work is organised as follows. Section \ref{Sec2} contains the theory and methodology used throughout this work. Subsection \ref{Sec21} elaborates on the general molecular Hamiltonian, and gives insight into the approximations and terminology used to study non-adiabatic couplings. Subsection \ref{Sec22} focuses on the electronic interactions in the ULRM and introduces the electronic states and PECs relevant to our work, before we discuss the specific two-level system and the problem of diabatization in subsections \ref{Sec23} and \ref{Sec24}, respectively. The computational tools used as well as the approach to obtain the vibronic spectra are discussed in section \ref{Sec3}, whereas section \ref{Sec4} features the results and discussion. Section \ref{Sec4} is further divided into subsections, with \ref{Sec41} addressing the electronic structure of the $^{23}$Na ULRM, \ref{Sec42} discussing the non-adiabatic couplings and \ref{Sec43} providing a comparative analysis of the coupled-channel vibronic spectrum, and the single-channel approximations. Our conclusions and outlook are presented in Section \ref{Sec5}.

\section{Theory and Methodology}    \label{Sec2}

In this section, we present the Hamiltonian governing the internal molecular dynamics of the ULRM, and the corresponding Schrödinger equation in atomic units. We then proceed to elaborate on the non-adiabatic features of our system and the approximations used to circumvent them. Later, we introduce the scattering interaction which forms the binding mechanism of the ULRM, and discuss the electronic spectra of the molecule. The focus is then shifted to the interaction between the trilobite and the $d$-state, and the vibronic coupling between them. Finally, we introduce the concept of diabatization and use it in our two-state system as a precursor to obtaining the vibronic spectra.

\subsection{Hamiltonian, adiabatic separation and non-adiabatic couplings}    \label{Sec21}

Our system consists of two $^{23}$Na atoms, one in the Rydberg state and the other one in the ground state. Transforming the two-atom system into relative coordinates, $\textbf{R}$ is the internuclear vector of the two atoms, and $\textbf{r}$ is the Rydberg electron position, with the ionic core at the coordinate origin, as depicted in figure (\ref{fig1}). The Rydberg electrons' interaction with the parent core and the ground-state perturber, as well as the vibrational motion of the diatomic system are captured in the molecular single-electron effective Hamiltonian:

\begin{figure}
    \includegraphics[width=0.38\textwidth]{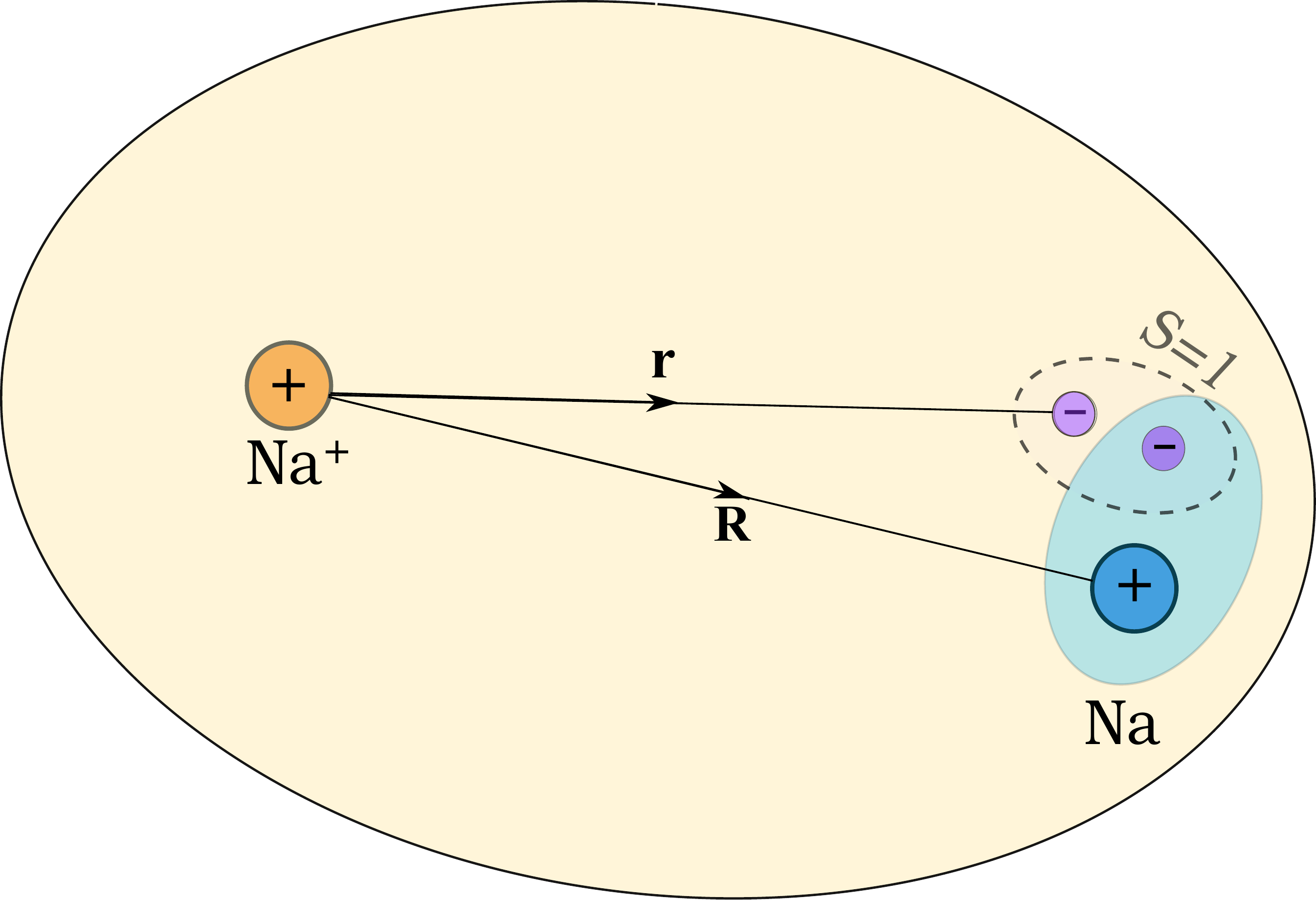}
    \caption{Sketch of the $^{23}$Na ULRM, illustrating the scattering interaction between the Rydberg electron at position $\textbf{r}$ and the ground state atom at position $\textbf{R}$. The triplet nature of the scattering interaction is highlighted. }
    \label{fig1}
\end{figure}

\begin{equation}    \label{1}  
   \textbf{H}_{\mathrm{m}} =  \frac{\textbf{P}^2_{\mathrm{nu}}}{2\mu} + \underbrace{\frac{\textbf{p}^2_\mathrm{e}}{2\mu_{\mathrm{e}}} + V_{\mathrm{Ryd}} (\textbf{r}) + V_{\mathrm{en}}(\textbf{R},\textbf{r})}_{\textbf{H}_{\mathrm{e}}(\textbf{R},\textbf{r})}
\end{equation}
\newline
The first term represents the nuclear kinetic energy $T_{\mathrm{nu}}$ along the vibrational degree of freedom, the second term and the third term represent the kinetic energy of the Rydberg electron $T_{\mathrm{e}}$ and its interaction with the Rydberg ionic core $V_{\mathrm{Ryd}}$, whereas the last term $V_{\mathrm{en}}$ represents the electron-perturber interaction. $\mu$ and $\mu_{\mathrm{e}}$ are the reduced masses of the two nuclei and the electron, respectively. Isolating $\textbf{H}_{\mathrm{e}}(\textbf{R},\textbf{r})$ as the electronic Hamiltonian which is parametrically dependent on the internuclear coordinate, one could solve for the corresponding electronic problem. The resulting eigenvalues, i.e.~the potential energy curves (PECs) $\varepsilon_i(\textbf{R})$ and eigenvectors $\psi_i(\textbf{R},\textbf{r})$ depend parametrically on the internuclear coordinates, and we have

\begin{equation}    \label{2}
  \textbf{H}_{\mathrm{e}}(\textbf{R},\textbf{r}) \ket{\psi_i(\textbf{R},\textbf{r})} = \varepsilon_i(\textbf{R}) \ket{\psi_i(\textbf{R},\textbf{r})}.
\end{equation}
 \newline
 The total molecular wavefunction can be expanded using the electronic eigenfunctions according to 
 $\Psi_{\mathrm{m}} (\textbf{R},\textbf{r}) = \sum_i \chi_i(\textbf{R}) \psi_i(\textbf{R},\textbf{r})$, known as the Born-Oppenheimer expansion \cite{Born_1927,Huang_1954}. It is an exact representation of the molecular wavefunction, as the electronic eigenfunctions form an orthonormal and complete basis. Here $\{\chi_i(\textbf{R})\}$ are the expansion coefficients which portray the $\textbf{R}$-dependent mixing of electronic states. Inserting this expansion into the molecular Schrödinger equation $\textbf{H}_{\mathrm{m}}\Psi_{\mathrm{m}} = E \Psi_{\mathrm{m}}$, and integrating out the electronic degrees of freedom, we obtain the coupled-channel time-independent Schrödinger equation (TISE) for the vibrational motion \cite{koppel_1984}:
 
\begin{equation}  \label{3}
  \begin{split}
       -\frac{1}{2\mu}\bm{\nabla}_\textbf{R}^2 \chi_i(\textbf{R}) +  \varepsilon_i(\textbf{R}) \chi_i(\textbf{R}) &-  \frac{1}{2\mu}\sum\limits_j \Lambda_{ij}(\textbf{R}) \chi_j(\textbf{R}) \\ &= E \chi_i(\textbf{R}).
  \end{split}     
\end{equation}
\newline
Here, $\Lambda_{ij}$ are the non-adiabatic couplings between the nuclear and electronic motions, which can be written as 
\begin{equation}    \label{4}
    \Lambda_{ij} = 2 P_{ij} \bm{\nabla}_\textbf{R} + Q_{ij},
\end{equation}
\newline
where $P=(P_{ij})$ and $Q=(Q_{ij})$ are the first and second order derivative couplings, defined as:
\begin{equation}    \label{5}
    P_{ij} =  \braket{\psi_i(\textbf{R},\textbf{r})|\bm{\nabla}_\textbf{R}|\psi_j(\textbf{R},\textbf{r})},
\end{equation}
\begin{equation}    \label{6}
    Q_{ij} =  \braket{\psi_i(\textbf{R},\textbf{r})|\bm{\nabla}^2_\textbf{R}|\psi_j(\textbf{R},\textbf{r})},
\end{equation}
\newline
where $\braket{}$ denote the integration w.r.t.~the electronic degrees of freedom. $P$ and $Q$ essentially introduce the coupling between different electronic states $\psi_i$ and $\psi_j$ due to the motion of the nuclei, which in turn affects the vibrational motion of the nuclei. Once the $P$ matrix is obtained, it is straight forward to calculate the second order coupling using the relation,
\begin{equation}    \label{7}
    Q_{ij} = \bm{\nabla}_\textbf{R} P_{ij} + P^2_{ij},
\end{equation}
\newline where, 
\begin{equation}    \label{8}
    P^2_{ij} = - \braket{\bm{\nabla}_\textbf{R} \psi_i(\textbf{R},\textbf{r}) | \bm{\nabla}_\textbf{R} \psi_j(\textbf{R},\textbf{r})}.
\end{equation} It is possible to write equation (\ref{3}) in a more compact way, using the complete $P$ matrix as:

\begin{equation}    \label{9}
    -\frac{1}{2\mu}(\bm{\nabla}_\textbf{R} + P)^2 \bm{\chi}(\textbf{R})  + \varepsilon(\textbf{R})\bm{\chi}(\textbf{R}) = E\bm{\chi}(\textbf{R}),
\end{equation}
\newline
with the vector $\bm{\chi} = (\chi_i)$, that contains the expansion coefficients, and the diagonal potential energy matrix $\varepsilon$, that contains the corresponding PECs. The non-adiabatic couplings are shown to manifest as off-diagonal terms in the kinetic energy operator for the nuclear vibrational motion.

The Born-Oppenheimer (BO) approximation \cite{Born_1927,koppel_1984,Huang_1954,Agostini_2019,Baer_2006} is used to solve equation (\ref{3}) as a standard approximation, where the non-adiabatic terms are completely neglected, resulting in the decoupling of the vibrational Schrödinger equations governing $\chi_i$ to form:

\begin{equation}     \label{10}
    -\frac{1}{2\mu}\bm{\nabla}_\textbf{R}^2 \chi_i(\textbf{R}) +  \varepsilon_i(\textbf{R})\chi_i(\textbf{R}) = E_i \chi_i(\textbf{R}).
\end{equation}
\newline
Here, the total molecular state is simply, $\Psi_{\mathrm{m},i} (\textbf{R},\textbf{r})=\chi_i(\textbf{R}) \psi_i(\textbf{R},\textbf{r})$, i.e.~the electronic and nuclear motion are adiabatically separated. $\psi_i(\textbf{R},\textbf{r})$ is the adiabatic eigenstate and $\varepsilon_i(\textbf{R})$ is the adiabatic PEC which describe the electronic motion, and $\chi_i(\textbf{R})$ is the vibrational wavefunction on each PEC. These adiabatic states and energy surfaces can be obtained by solving $\textbf{H}_{\mathrm{e}}$ at each fixed nuclear geometry. Equation (\ref{10}) then describes the vibrational motion in each individual adiabatic potential energy curve. The adiabatic approximation can be justified due to the large differences in masses between the electrons and the nuclei, resulting in very different time-scales of their motions. Hence the nuclei are approximated to remain frozen, over the course of the electronic dynamics.

In a second approximation, called the Born-Huang (BH) approximation \cite{Huang_1954}, we include the diagonal derivative coupling operator $\Lambda_{ii}$ in equation (\ref{10}), 

\begin{equation}    \label{11}
    -\frac{1}{2\mu}\bm{\nabla}_\textbf{R}^2 \chi_i(\textbf{R}) +  \varepsilon^b_i(\textbf{R})\chi_i(\textbf{R}) = E_i \chi_i(\textbf{R}),
\end{equation}
\newline
where,
\begin{equation}    \label{12}
    \varepsilon^b_i(\textbf{R}) = \varepsilon_i(\textbf{R}) - \frac{1}{2\mu}\Lambda_{ii}(\textbf{R}).
\end{equation}
The Born-Huang approximation still maintains the decoupling between different adiabatic electronic states due to nuclear motion, instead it merely adds a correction to each isolated PEC due to the finite kinetic energy of the nuclei. Hence the BH approximation, much like the BO approximation results in a single-channel TISE by ignoring the off-diagonal couplings of the coupled-channel TISE, equation (\ref{3}). Note that $P$ is anti-hermitian, whereas $Q$ is non-hermitian. Hence, if the electronic eigenfunctions are real, $P_{ii}$, the diagonal first order term vanishes and we obtain $(\Lambda_{ii} = Q_{ii}) \leq 0$ (from equation(\ref{7},\ref{8})), thereby causing a purely positive shift in the adiabatic PEC.

\begin{figure*}
    \centering
    \centerline{\includegraphics[width=0.9\textwidth]{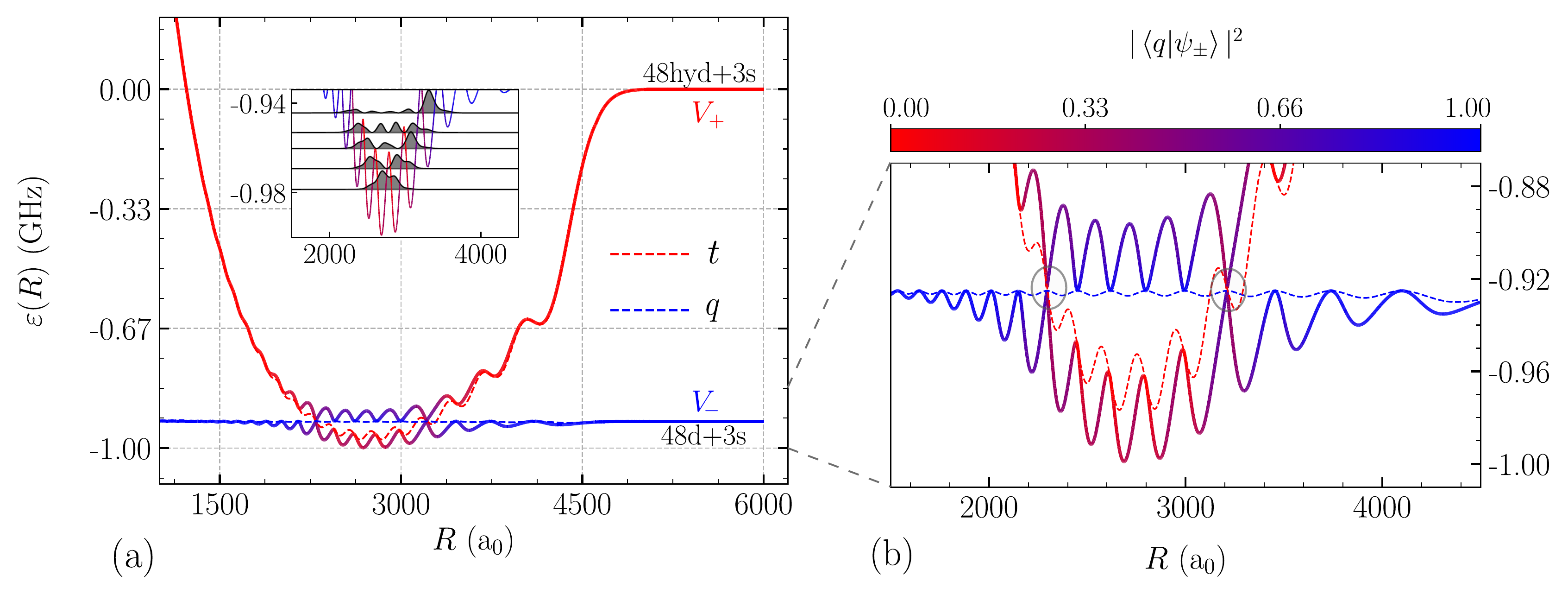}}
    \caption{(a) Adiabatic and diabatic potential energy curves for $n$=48. The $ V_{-}$ and $t$ PECs converge to the $48d$ electronic state as it approaches the dissociation limit ($\sim$ 4500 $\mathrm{a}_0$) and  $ V_{+}$ and $q$ PECs converges to the $n$=48 hydrogenic state. The inset presents the lowest five vibrational levels in the $V_{-}$ PEC and their probability densities.
    (b)  Magnified view of the region with prevalent state mixing ($2000\leq R \leq 4000$ $\mathrm{a}_0$). The diabatic curves are allowed to cross each other, but the adiabatic curves show nearly degenerate avoided crossings at $R \sim 2295$ and 3210 $\mathrm{a}_0$ (encircled regions), respectively. The colorbar features the $l=2$ character of the PECs as a function of $R$.}
    \label{fig2}
\end{figure*}

Both the Born-Oppenheimer and the Born-Huang approximation are only valid when the $\textbf{R}$ dependence of $\psi_i$ and $\varepsilon_i$ are adiabatic, i.e.~the change of electronic motion with respect to $\textbf{R}$ is gradual. But both of these assumptions are broken in the vicinity of an avoided level crossing of two PECs, where the coupling between the electronic states are non-negligible due to a high $\textbf{R}$ sensitivity of the electronic states. The strong vibronic coupling effects between the vibrational and electronic states, due to the avoided crossings, can change the spectra and lifetimes of molecules and facilitate surface hopping across the avoided crossing.

\subsection{Electronic Interaction} \label{Sec22}

Solving the molecular Hamiltonian for a ULRM, first requires the characterization of the electronic interaction between the Rydberg electron and ground state atom, which perturbs the Rydberg electron wave-function via low-energy scattering. We model $V_{en}$, the interaction potential, using a Fermi pseudo potential \cite{Fermi1934,Omont_1977} for s-wave scattering:

\begin{equation}    \label{13}
    V_{en}(\textbf{r},\textbf{R}) = 2\pi a_s^T(k) \delta(\textbf{r}-\textbf{R}),
\end{equation}
\newline
where $a_s^T(k) = a_s^T(0) + \pi \alpha k/3$ is the energy-dependent triplet s-wave scattering length for electron collisions with ground state atoms, where $a_s^T(0) = -$5.9 is the zero energy triplet scattering length \cite{Eiles_2018,Karule_1965}, and $\alpha$ = 162.7 is the polarizability \cite{Holmgren_2010,Mitroy_2010,Ekstrom_1995}. The wave-number $k$, can be semi-classically determined using $k^2 = 2/R - 1/n^2$. The negative triplet scattering length facilitates an attractive interaction capable of binding the two $^{23}$Na atoms. A pictorial representation of the interaction is presented in figure (\ref{fig1}).

The adiabatic electronic Hamiltonian for the ULRM is hence given by, $H_{\mathrm{e}} = H_{\mathrm{Ryd}} + V_{\mathrm{en}}$, where $H_{\mathrm{Ryd}}$ is the Rydberg electron Hamiltonian. $H_{\mathrm{e}}$ can be diagonalized in the Rydberg-state basis $\{ \braket{\textbf{r}|nlm} = \phi_{nlm} (\textbf{r}) \}$, which satisfy the Rydberg atom's TISE,

\begin{equation}    \label{14}
    H_{\mathrm{Ryd}}\ket{nlm} = -\frac{1}{2(n-\mu_l)^2}\ket{nlm},
\end{equation}
\newline
where $\ket{nlm}$ is an atomic Rydberg eigenstate. In our case $\mu_l$ is the $l$-dependent quantum-defect of $^{23}$Na \cite{Lorenzen_1983,Hummel2021}, which decide the detuning of each quantum-defect state with reference to the hydrogenic state with energy $E_{n} = -1/2n^2$. The spherical symmetry of the Hamiltonian implies that, without loss of generality, the internuclear axis can be taken as the z-axis, and only $m=0$ states contribute. The high angular momentum states ($n-1\geq l\geq 3$ for Na) are approximately degenerate to the hydrogenic manifold due to negligible quantum-defects. Degenerate perturbation theory is used to obtain the high-$l$ adiabatic electronic states of $H_{\mathrm{e}}$,

\begin{equation}    \label{15}
    \ket{t} = \frac{1}{T}\sum_{l=3}^{n-1}  \phi_{nl0}(R\hat{\mathbf{z}}) \phi_{nl0}(\mathbf{r}),
\end{equation}
\newline
where,
\begin{equation}    \label{16}
    T^2 = \sum_{l=3}^{n-1} |\phi_{nl0}(R\hat{\textbf{z}})|^{2},
\end{equation}
\newline
and the corresponding potential energy curve,
\begin{equation}    \label{17}
    t(R) = 2 \pi a_{s}^T(k) \hspace{1mm} T^{2},
\end{equation}
\newline
which are popularly known as the trilobite state and trilobite PEC \cite{Greene2000,Hummel2020,Greene_1987}. The low angular momentum Rydberg-electron states ($l<3$), are energetically far-detuned from the hydrogenic manifold due to significant quantum-defects. Using non-degenerate perturbation theory, we obtain the isolated, low-$l$ adiabatic electronic state of $H_{\mathrm{e}}$,

\begin{equation}    \label{18}
    \ket{q} = \phi_{\nu l0}(\textbf{r}),
\end{equation}
and the corresponding potential energy curve
\begin{equation}    \label{19}
  q(R) = 2\pi a_{s}^T(k)|\phi_{\nu l0}(R\hat{\textbf{z}})|^{2},
\end{equation}
\newline
where $\nu = n-\mu_l$ is the effective principal quantum number. Note that the low-$l$ molecular electronic state is essentially the Rydberg-electron state, without any $R$-dependent state mixing. These two PECs, first introduced in 2000 \cite{Greene2000}, are the building blocks of the Rydberg ULRM. The trilobite PECs feature wells which are significantly deeper ($\sim$GHz) than their low-$l$ counterparts ($\sim$MHz) and can support multiple bound molecular states. These trilobite molecules are extremely polar with a large electric dipole moment $D \approx R -n^2/2$ $e \mathrm{a}_0$ (a$_0$ is the length scale in atomic units), as compared to the non-polar low-$l$ molecules, which can only support a few weakly bound states. This simple perturbative method covers the fundamental properties of Rydberg molecules, it allows us to study the two distinct classes of PECs and analyze the properties of both polar and non-polar molecules formed by them.

\subsection{Two Level System}   \label{Sec23}

As a prototype setup, the non-adiabatic interactions in $^{23}$Na ULRM, can be effectively probed by studying the interaction between the trilobite and $l=2$ i.e.~$d$ quantum-defect state relevant in sodium. We expand $H_{\mathrm{e}}$, in a restricted basis $\{\ket{t},\ket{q} \}$ consisting of the trilobite and $d$-state to obtain,

\begin{equation}    \label{20}
   V_{\mathrm{e}}= \begin{bmatrix}
\braket{t|H_{\mathrm{e}}|t} & \braket{q|H_{\mathrm{e}}|t} \\
\braket{t|H_{\mathrm{e}}|q} & \braket{q|H_{\mathrm{e}}|q}
\end{bmatrix}=  \hspace{1mm}\begin{bmatrix}
t & \sqrt{qt} \\
\sqrt{qt} & q + \Delta
\end{bmatrix},
\end{equation}
\newline
where $\Delta$ is the energy splitting between the the $d$-state with reference to the hydrogenic manifold \cite{Hummel2021}. Two adiabatic PECs are obtained from the diagonalization
\begin{equation}    \label{21}
    V_{\pm}(R) = \frac{1}{2} [ t+q+\Delta  \pm \sqrt{(t-q-\Delta)^{2} +4qt}  ],    
\end{equation}
\newline
along with their corresponding adiabatic electronic states,
\begin{equation}    \label{22}
\begin{split}
    \ket{\psi_{-}} &= \cos{\theta} \ket{t} + \sin{\theta} \ket{q}, \\
    \ket{\psi_{+}} &= -\sin{\theta} \ket{t} + \cos{\theta} \ket{q},
\end{split}
    \end{equation}
\newline
where $\theta$ is the mixing angle given by,
\begin{equation}     \label{23}
    \theta = \arccos(\frac{\sqrt{qt}}{\sqrt{(V_{-}-t)^2 + qt}}).
\end{equation}
\newline
The state mixing is largely determined by the detuning and the coupling strength. For $\Delta \gg q,t$, such that $\sqrt{qt}/(t-q-\Delta) \sim 0$, as is the case for $l<2$, we obtain $V_{\pm} \approx \{q, t\}$. For large detuning the adiabatic PECs act similar to the $l=2$ i.e.~$d$-state PEC and trilobite PEC independently. Here, we assume that the rotational angular momentum of the molecule is zero, thereby restricting the internuclear motion to the radial dimension, and establishing our system as a two-level single-parameter model. From this two-level picture, the derivative couplings between the two adiabatic states are determined to be

\begin{equation}    \label{24}
\begin{split}
    P_{12} &=  \braket{\psi_{-}|\partial_R|\psi_{+}} = -\theta', \\
    Q_{11} &= \braket{\psi_{-}|\partial_R^2|\psi_{-}} = - \theta'^2 + \cos^2(\theta)  \braket{t|\partial_R^2|t}, \\
    Q_{22} &= \braket{\psi_{-}|\partial_R^2|\psi_{-}} = - \theta'^2 + \sin^2(\theta)  \braket{t|\partial_R^2|t},
\end{split}
\end{equation}
\newline
where,
\begin{equation}    \label{25}
    \braket{t|\partial_R^2|t} = \frac{M^4}{T^4} - \frac{B^2}{T^2},
\end{equation}
\newline
with$\hspace{2mm} \theta'$=$\partial_R \theta,\hspace{2mm} B^2$=$\sum_l |\phi'_{nl0}(R)|^2, \hspace{2mm}$and$\hspace{2mm} M^2$= $\sum_l \phi'_{nl0}(R) \phi_{nl0}(R)$.
\newline

\begin{figure}
    \centering
    \includegraphics[width=0.45\textwidth]{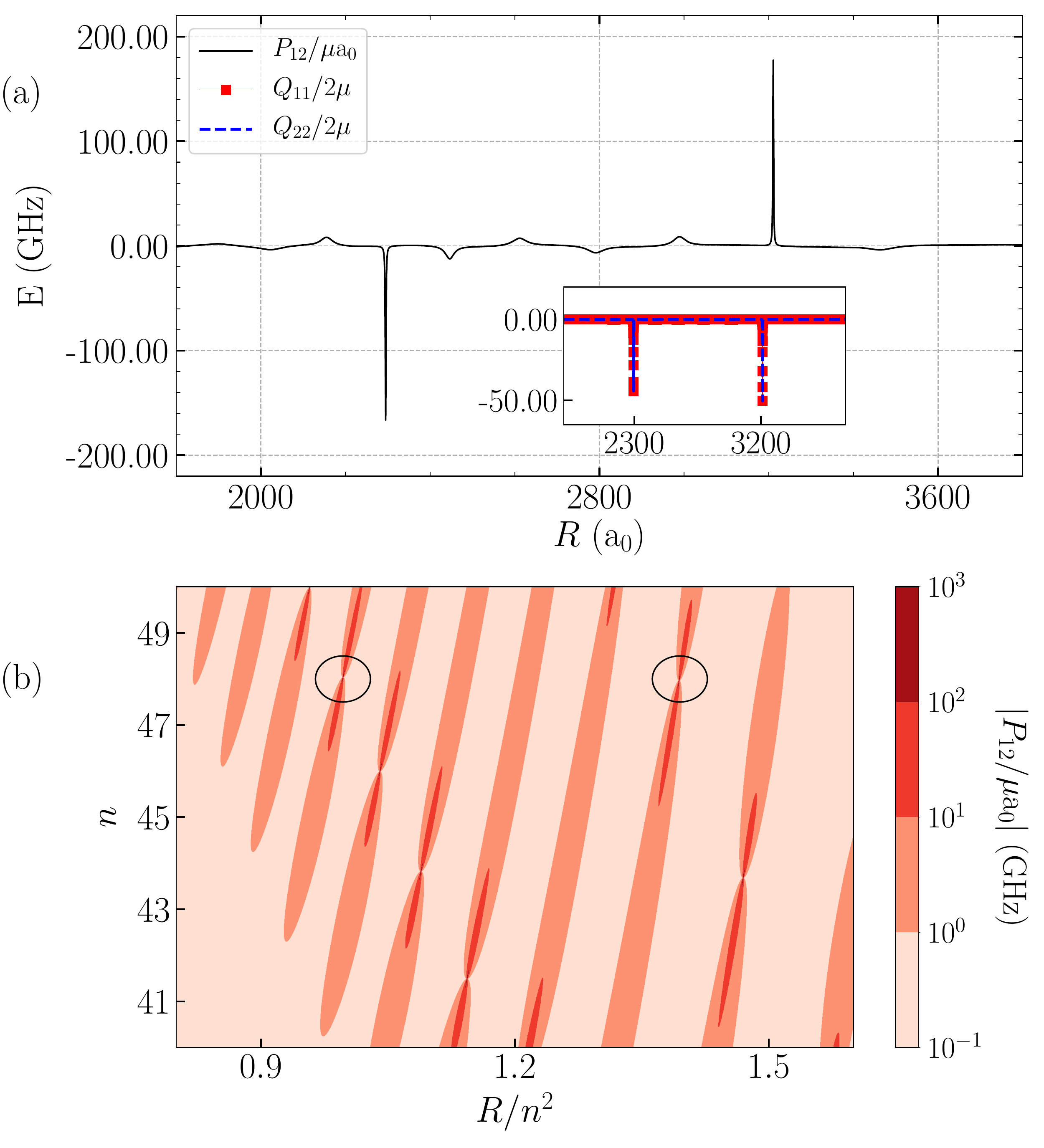}
    \caption{The non-adiabatic couplings of the adiabatic electronic states $\ket{\psi_{-}}$ and $\ket{\psi_{+}}$. a) The first order derivative coupling (black) $P_{12}/\mu$, for $n$=48.  The inset depicts the second order derivative coupling $Q_{11}/2\mu$ (red) and $Q_{22}/2\mu$ (blue). Both the first and second order derivative couplings portray two near-singular peaks corresponding to the avoided level crossings at $R \sim 2295$ and 3210 $\mathrm{a}_0$, respectively. b) Parameter scan of the magnitude of $P_{12}(n,R)$ over varying $n$. Multiple $n$ values are found possessing strong non-adiabatic couplings, the peaks near $n$=48 are encircled. }
    \label{fig3}
\end{figure}

Figure (\ref{fig3}a) shows the calculated derivative couplings for $n$=48. The derivative couplings determined are prone to be divergent for certain $n$ and $R$. The existence of these singularities \cite{Worth_2004} pose numerical problems in solving equation (\ref{3}), hence it is of use to employ methods whereby the vibrational Schrödinger equation is solved without dealing with divergent derivative couplings.

\subsection{Diabatization}  \label{Sec24}

The derivative coupling terms, as discussed before, manifest as off-diagonal terms based on the nuclear kinetic energy operator. The idea of diabatization \cite{koppel_1984,Baer_2006} is to perform a unitary transformation which diagonalizes the kinetic energy matrix, whereby the need for derivative couplings in solving the vibrational TISE is circumvented. As a result of such a transform, off diagonal potential energy terms may be introduced in the new basis, which accounts for the vibronic couplings in the diabatic basis. A unitary transformation $\bm{\Bar{\chi}}(R) = U(R) \bm{\chi}(R)$, would transform the TISE (equation (\ref{9})) to,

\begin{equation}    \label{26}
     -\frac{1}{2\mu}(\mathbb{I} \hspace{0.5mm} \partial_R + \bar{P})^2 \bm{\Bar{\chi}}(R)  + \Bar{\varepsilon}(R)\bm{\Bar{\chi}}(R) = E\bm{\Bar{\chi}}(R),
\end{equation}
\newline
where,
\begin{equation}     \label{27}
    \begin{split}
        \Bar{\varepsilon}(R) = U^{\dagger} \varepsilon (R)U  , \hspace{5mm}
        \bar{P} = U^{\dagger} ( P U  + \partial_R U).
    \end{split}
\end{equation}
The transformed derivative coupling matrix vanishes if ,
\begin{equation}    \label{28}
    \partial_R U = -P U,
\end{equation}
\newline
in which case the matrix $U$ represents the diabatic transform, and $\Bar{\varepsilon}$ is the diabatic potential energy matrix with off-diagonal terms. While it may not be possible to construct such a unitary transform in all situations, for a two level system with a single parameter R, such a transformation exists analytically \cite{Top_1975,hollerith_2019} where $U$ is a rotation about the angle,

\begin{equation}    \label{29}
    \gamma(R) = \int_{R}^{\infty} P_{12}(R_1)dR_1.
\end{equation}
\newline
 Using the corresponding form of $P_{12}$, we obtain $\gamma(R) = -\int_{R}^{\infty} \theta'(R_1)dR_1 = -\theta(R) $. Hence, the diabatic basis is proven to be the restricted electronic basis $\{ \ket{t},\ket{q} \}$, which was used as the initial ansatz. This gives the diabatic potential matrix

\begin{equation}    \label{30}
   \bar{\varepsilon} \hspace{1mm} =  \hspace{1mm}\begin{bmatrix}
t & \sqrt{qt} \\
\sqrt{qt} & q + \Delta
\end{bmatrix},
\end{equation}
  What is left is to obtain the vibronic spectrum, by solving the coupled vibrational Schrödinger equations, using the given diabatic potential energy matrix, the elements of which vary as smooth functions of $R$, devoid of any singularities.

\section{Computational Approach: Vibronic structure}    \label{Sec3}

Based on the diabatic PECs we use a tenth order finite difference method \cite{Groenenboom_1990} to obtain the vibrational wave-functions by solving the TISE for each adiabatic PEC. Convergent eigenvalues and wavefunctions were observed for a grid of (1000-6000) $ \mathrm{a}_0$, with a step-size of $1 \mathrm{a}_0$. Convergence of $P$ and $Q$ demands much finer grid steps near the avoided crossings.  This is overcome by diabatization; using the smooth diabatic PECs, we obtain a convergent eigenspectrum for the coupled-channel vibronic system for a step-size of $1  \mathrm{a}_0$. Once the diabatic basis is obtained, the coupled-channel vibronic Hamiltonian in the diabatic basis can be represented as,

\begin{equation}    \label{31}
    H_{\mathrm{m}}=\begin{bmatrix}
        T_{\mathrm{nu}} + t(R) & \sqrt{qt} (R) \\
        \sqrt{qt} (R) & T_{\mathrm{nu}} + q(R) + \Delta
    \end{bmatrix},
\end{equation}
\newline
 where $T_{\mathrm{nu}}$ is the nuclear kinetic energy term acting on each of the diagonal PECs. Solving the TISE $H_{t} = T_{\mathrm{nu}} + t(R)$, $H_{q} = T_{\mathrm{nu}} + q(R)$ on the aforementioned grid gives us the eigenspectrum $\{E_i^{t}, \ket{\chi_i^{t}}\}$, $\{E_i^{q}, \ket{\chi_i^{q}}\}$ corresponding to each diagonal diabatic PEC. We can then expand any eigenvector of $H_{\mathrm{m}}$ as
 \begin{equation}    \label{32}
        \ket{\Psi_{\mathrm{m}}^j} = \sum_{i=0}^{N-1} c_i^j \ket{\Bar{\chi}_i^t} \ket{t} +  \sum_{i=0}^{N-1} d_i^j \ket{\Bar{\chi}_i^q} \ket{q},
 \end{equation}
 \newline
 where,
  \begin{equation}    \label{33}
        H_{\mathrm{m}}\ket{\Psi_{\mathrm{m}}^j} = E^j\ket{\Psi_{\mathrm{m}}^j}.
 \end{equation}
 \newline
Here $\{c^j\}$ and $\{d^j\}$ are expansion coefficients obtained by diagonalizing $H_{\mathrm{m}}$ in the diabatic basis, and N is the number of vibrational states from each diagonal PEC used for diagonalization. Note that the eigenvector can no longer be separated into well-defined electronic or vibrational parts. However, tracing out the electronic degree of freedom leaves us with the vibrational probability density of $\ket{\Psi_{\mathrm{m}}^j}$ along the internuclear axis,
 
 \begin{equation}   \label{34}
     P^j(R) = |\Bar{X}^j_t(R)|^2 + |\Bar{X}^j_q(R)|^2,
 \end{equation}
\newline
 where,
 \begin{equation}   \label{35}
     \Bar{X}^j_t(R) = \sum_{i=0}^{N-1} c_i^j \Bar{\chi}_i^t(R), \hspace{5mm} 
     \Bar{X}^j_q(R) = \sum_{i=0}^{N-1} d_i^j \Bar{\chi}_i^q(R).
 \end{equation}
 \newline
 $P^j(R)$ can be compared to the probability density of the vibrational states in each adiabatic PEC (figure (\ref{fig4},\ref{fig5})), whereby differences in vibrational motion can be observed between the adiabatic and non-adiabatic case. To observe electronic state mixing, we transform the vibronic eigenstate back to the adiabatic basis where $\bm{X}^j= U^{\dagger}\bm{\Bar{X}}^j$, where $U^{\dagger}$ is the inverse-diabatic transform, to obtain
 
 \begin{equation}   \label{36}
     \ket{\Psi_{\mathrm{m}}^j} = X^j_{-}(R) \ket{\psi_{-}} + X^j_{+}(R)\ket{\psi_{+}},
 \end{equation}
 \newline
 where $X_{\pm}$ provides the $R$-dependent electronic-state mixing due to the non-adiabatic couplings, and
\begin{equation}    \label{37}
    p_{\pm} = \int |X_{\pm}(R)|^2 dR,
\end{equation}
 \newline
 which provide the population for the state $\ket{\psi_{\pm}}$. Color-mapping the $P^j(R)$ with the population $p_{-}$ (see figures (\ref{fig4},\ref{fig5})) is an effective way to visualize both the electronic and vibrational contribution to each vibronic eigenstate. The unbound states above the dissociation threshold suffer box-state behavior due to the fixed boundary conditions we impose, and are sensitive to the changes in the boundary-wall position. Hence, only the bound states below the threshold, and resonance states above threshold are considered in our analysis. We ensured convergence of these states with respect to variation of both boundary position and grid-size.

 \section{Results and Discussion}   \label{Sec4}
 
 In this section, we discuss the electronic and vibronic spectral properties of the sodium ULRM. The principal quantum number $n=$48 is used to illustrate the structure of the adiabatic PECs and derivative coupling terms, as it is representative for strong vibronic interaction. We then proceed to discuss the complete vibronic spectra in comparison to the BO and the BH spectra and elaborate on the specific differences between them due to non-adiabatic interaction effects. The vibronic spectrum for $n=$43 is also discussed, as it features less prominent vibronic coupling when compared to the $n=$48 case.
 
\subsection{Electronic structure}   \label{Sec41}

Figure (\ref{fig2}) features the two-state adiabatic PECs, $V_{\pm}$ determined for $n$=48 (solid lines). At large $R$, close to the dissociation limit ($R\sim$ 4500 a$_{0}$), the PECs become nearly flat and match the energies of the hydrogenic manifold ($V_{+}$) and the $d$-state ($V_{-}$), respectively. Inside the Rydberg orbit, $R \leq 2n^2$, the $\ket{\psi_+}$ electronic state splits off of the hydrogenic manifold, causing the $V_{+}$ PEC to descend towards the $V_{-}$ PEC and form narrow avoided crossings at $R \approx$ \{2295, 3210\} $\mathrm{a}_0$, where non-adiabatic couplings become large. 

The exchange of the electronic state character due to state mixing at avoided crossings is given by color-grading the adiabatic PECs, with the $R$ dependent overlap of the adiabatic states with the $d$-state. Corresponding to the mixing of electronic states, the $V_{\pm}$ PECs split away from $q(R)$ and $t(R)$ for $1700 \leq R \leq 4300$ $\mathrm{a}_0$. For $R$ values between the avoided crossings, $V_{-}$ acquires a trilobite character, as is visible from the well structure of the PEC, and the low $d$-state contribution to $\ket{\psi_{-}}$. $V_{+}$ on the other hand, while being $d$-state dominant, has a significantly altered well-structure. Note that, although $q(R)$ appears to be flat in the relevant energy-scale, it does exhibit oscillatory structures of depth $\leq 6$~MHz, significantly different in depth and structure to $V_{-}$ for the complete $R$ interval within which potential wells appear. The mixing of electronic states leads to the modification of transition dipole moments, facilitating the excitation of high-$l$ states via two-photon transitions. We remark that the validity of our surfaces $V_{\pm}$ is confirmed by comparison to numerically obtained adiabatic PECs including atomic Rydberg states belonging to energetically neighbouring $n$ manifolds, as well as corresponding quantum defect splitted states.

Furthermore, the single-color dashed lines are used to represent the diabatic PECs, corresponding to the trilobite state (red) and the $d$ state (blue), respectively. Note that the diabatic curves cross each other at $R$ values corresponding to the avoided crossings. This is expected, as the crossing of diabatic curves, given by the relation $t = q+\Delta$, minimizes the energy gap between the adiabatic PECs to $|V_{+}-V_{-}| = \sqrt{4qt}$ (see equation (\ref{21})). Due to the oscillatory nature of the $d$-state PEC, it is possible that the diabatic curves cross each other near the node of the $\ket{nd0}$ eigenstate. Such a coincidence would result in an extremely narrow avoided crossing, as is illustrated in figure (\ref{fig2}) for the special case of $n$=48.

\begin{figure*}
    \includegraphics[width=\textwidth]{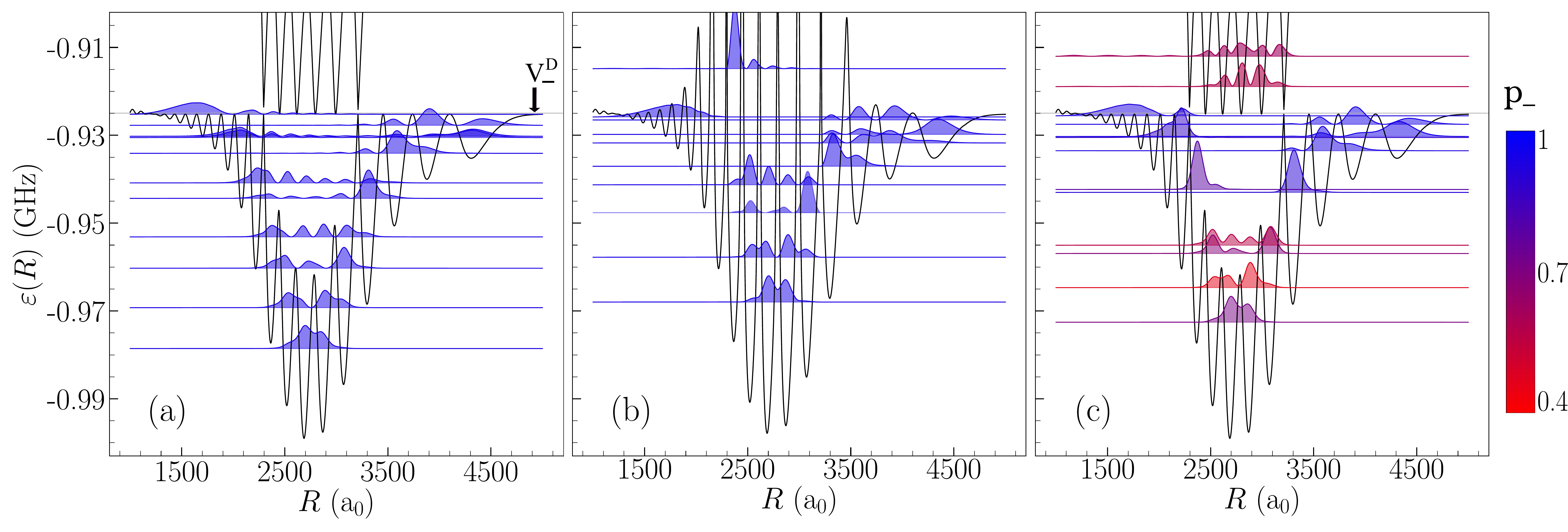}
    \caption{The eigenspectrum of the ultralong-range sodium molecule for $n$=48. Selected vibrational energy-levels and probability densities of bound states corresponding to the (a) $V_{-}$ Born-Oppenheimer curve (b) $V^b_{-}$ Born-Huang curve and (c) the coupled-channel non-adiabatic system, with the adiabatic curves plotted for reference. Each vibrational level is color-graded according to the electronic contribution by the $\ket{\psi_{-}}$ state, represented by the population $p_{-}$. The dissociation threshold of the $V_{-}$ PEC, given by $V^D_{-} = -0.925$ GHz is presented in sub-figure (a).}
    \label{fig4}
\end{figure*}

Extending our analysis, if the $t = q+\Delta$ crossing exactly coincides with the node of $\ket{nd0}$, then $q$ vanishes and the adiabatic PECs become degenerate, i.e.~$|V_{+}-V_{-}|=0$. However, the von Neumann-Wigner non-crossing theorem prohibits an exact degeneracy of the PECs corresponding to electronic states of the same symmetry, in single-parameter (determined by $R$) diatomic systems \cite{vonNeumann_1993}. We note that the use of the principal quantum number $n$, as a synthetic dimension \cite{Hummel2021}, allows us to bypass the non-crossing theorem. This is facilitated by using Whittaker Coulomb functions as a replacement for the hydrogenic radial wavefunctions, for probing non-integer $n$ \cite{Eiles_2019,Wang_2020}. The two-parameter, diatomic system, introduces the possibility of forming conical intersections at specific $\{n,R\}$ coordinates. Such CIs would result in complete degeneracy of the adiabatic potentials, representing $\{n,R\}$ values where the diabatic curves cross each other at exactly the node of $\ket{nd0}$. Hence, the width of the avoided crossings and consequently the strength of the non-adiabatic coupling is ascertained to be $n$-dependent, corresponding to the proximity of the $\{n,R\}$ coordinate to a CI.

\subsection{Non-adiabatic coupling} \label{Sec42}

Figure (\ref{fig3}) features the non-adiabatic couplings of the two adiabatic electronic states of the ULRM. Since the $P$ matrix is anti-hermitian, it suffices to analyze $P_{12}$ (= $-P_{21}$) as off-diagonal coupling. $P_{12}$ is divided by the reduced mass $\mu$ and $\mathrm{a}_0$, so that it can be expressed in units of energy, as represented in equation (\ref{3}). Similarly, $Q_{11}$ and $Q_{22}$ are divided by twice the reduced mass. The non-adiabatic couplings are then expressed in the same energy scale as the adiabatic PECs, allowing to us to compare their relative strengths as they appear in the vibrational TISE (see equation (\ref{3})) and consequently establish their importance in the calculation of the eigenspectrum. 

 The derivative off-diagonal coupling for $n$=48 (see figure (\ref{fig3}a)), exhibits two near-singular peaks corresponding to the same $R$ values as the narrow avoided crossings of the underlying PECs. At other $R$ values, $P_{12}$ also exhibit oscillatory structures with smaller amplitude, which are caused by the well-structures of $V_{+}$ and $V_{-}$ resulting in an oscillatory potential energy difference between them. The couplings vanish beyond $R>4500$ $\mathrm{a}_0$ and $R>1500$ $\mathrm{a}_0$, due to a negligible $R$ dependent state mixing. The second order non-adiabatic couplings $Q_{11}$ and $Q_{22}$ are both dominated by the contribution from $P_{12}$, overpowering the derivative term in equation (\ref{7}). Hence both the diagonal terms feature near-singular peaks corresponding to the narrow avoided crossings, with negligible strength at other $R$-values, on the displayed scale. The magnitude of the peaks in the non-adiabatic couplings far exceed the energy scale of the adiabatic potential energy curves near the avoided crossing, thereby causing the breakdown of the BO approximation. The strong contribution from the off-diagonal term which dominates the diagonal term implies that the BH approximation is also not applicable. Note that the magnitude and double peak structure of the non-adiabatic couplings are specific to the $n$=48 case, which allow the extremely narrow avoided crossings.

To further study the $n$-dependency of non-adiabatic effects, we determine the magnitude of $P_{12}$ over a parameter range of the principal quantum number (see figure (\ref{fig3}b)). The linear structures formed by $P_{12}$ stem from the fact that the length scale of the ULRM and hence the position of the avoided crossings scale as $n^2$. As explained before, these avoided crossings turn into CIs when the trilobite PEC crosses the $d$-state PEC, exactly at the zero of $\ket{nd0}$. Figure (\ref{fig3}b) features near-singular peaks for specific $n$-values which corresponds to the existence of such conical intersections accounting for synthetic dimensions. The $n$=48 case is highlighted, as it is very close to two CIs, which explains the existence of two nearly degenerate avoided crossings and the double peak structure of non-adiabatic couplings. Other integer $n$-values which are not in the vicinity of a CI, also feature prominent but less peaked derivative couplings which corresponds to their respective avoided crossings, that are relatively wider compared to the $n$=48 case.

\subsection{Vibrational and Vibronic spectra} \label{Sec43}

We now perform a comparative analysis of the spectral features of the sodium ULRM, obtained using the BO approximation, the BH approximation, and the coupled-channel Hamiltonian for $n$=48. Figure (\ref{fig4} a,b) show the vibrational probablity densities of selected bound eigenstates shifted according to their energy eigenvalues calculated using the BO and BH approximations, respectively. Figure (\ref{fig4}c) shows the vibrational probability density of selected bound eigenstates of the coupled-channel Hamiltonian, obtained by tracing out the electronic degree of freedom, shifted according to their energy values. The probability densities are colored according $p_{-}$, the population of state $\ket{\psi_{-}}$. The black lines represent the $V_{-}$, $V_{+}$ PECs in figure (\ref{fig4} a,b), and the BH corrected $V^b_{-}$ PEC in figure (\ref{fig4}c).

We observe that the vibronic eigenvalues exhibit a significant positive shift in energy, when compared to the vibrational eigenvalues obtained using the BO approximation. For $n$=48, the full vibronic system possesses a ground state shifted $\sim$6~MHz higher in energy than the BO ground state, and exhibits it negligible energy shifts for bound states near the continuum threshold. The energy gaps between the eigenvalues of the BO system fails to mimic the energy gaps exhibited by the coupled-channel non-adiabatic system. The Born-Huang correction shifts the vibrational ground state $\sim$11~MHz higher in energy than the BO ground state, effectively placing it above the first excited state of the BO system and above the ground state of the coupled-channel system. The higher excited states near the continuum also show significant positive energy shifts ($\sim$4~MHz) between the BO and the BH approximations. The over correction in energy values, due to the BH correction, is a result of including only the positive diagonal terms without considering the level-repulsion introduced by the off-diagonal terms.

\begin{figure*}
    \includegraphics[width=\textwidth]{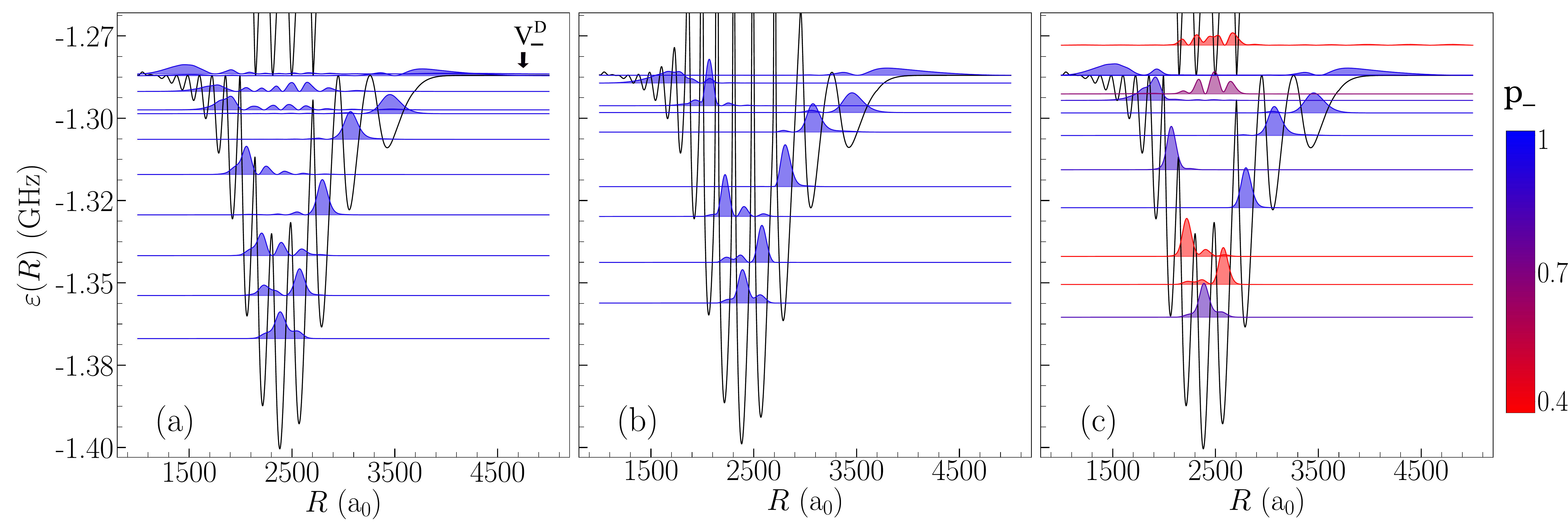}
    \caption{The eigenspectrum of the ULRM for $n$=43. Selected vibrational energy-levels and probability densities of bound states corresponding to the (a) $V_{-}$ Born-Oppenheimer curve (b) $V^b_{-}$ Born-Huang curve and (c) the coupled-channel non-adiabatic system, with the adiabatic curves plotted for reference. Each vibrational level is color-graded according to the electronic contribution by the $\ket{\psi_{-}}$ state, represented by the population $p_{-}$. The dissociation threshold of the $V_{-}$ PEC, given by $V^D_{-} = -1.287$ GHz is presented in sub-figure (a).}
    \label{fig5}
\end{figure*}

We also observe pronounced differences in the vibrational probability densities obtained within the three models. The vibrational ground state associated with the $V_{-}$ PEC is delocalized between several wells (2500 to 3300 $\mathrm{a}_0$). The ground state belonging to the BH and exact spectra, although similar in density profile, is more localized. Subsequent excited states belonging to the BH and exact spectra, while exhibiting different density profiles, also show more localized behavior. For example, the fourth excited BO state in figure (\ref{fig4}a) is delocalized in $R$ over the entire of the $V_{-}$ PEC at the relevant energy-scale. In contrast, the corresponding coupled states presented in figure (\ref{fig4}c), exhibits localization in $R$ over the span of a single potential well. Similar effects of localization is also visible clearly in the fifth and seventh excited states of the system. The higher potential barrier introduced between the potential wells, by the diagonal BH correction can be viewed as an explanation for vibrational localization in the BH spectrum. But the vibronic spectrum has a significantly altered probability distribution that is not explained by the diagonal correction alone. Note that below the continuum threshold of $-0.925$ GHz, the BO and BH spectra contains states with purely $\psi_{-}$ character. Counter-intuitively we observe significant state mixing for the first three excited states of the coupled system, even though they appear to be well localized in the $V_{-}$ PEC. This indicates strong non-adiabatic couplings between the two closed channels, well below the continuum limit.

A key observation to be made is the existence of non-adiabatic resonance states visible above the $V_{-}$ dissociation threshold. For energies above $-0.925$ GHz, the $n$=48 $V_{-}$ adiabatic PEC corresponds to an open channel, only featuring continuum states. However, in the complete vibronic picture (figure (\ref{fig4}c)), bound states are observed at energies $\sim$ {$-0.919, -0.912$ GHz}, with significant state-mixing (determined by color-value). The vibrational probability density of these bound states are localized in the potential energy wells of $V_{+}$, but are localized much below the ground state of $V_{+}$, hence, they cannot be interpreted to belong to either one of the single channels. The BH spectra does contain a bound state of energy $-0.915$ GHz localized in the $V^b_{-}$ PEC, but this does not compare in energy value or probability density to the aforementioned resonance states. The bound states are hence explained as scattering resonances due to the off-diagonal non-adiabatic coupling between an open channel (limited to $V_{-}$) and a closed channel (limited to $V_{+}$).

Figure (\ref{fig5}) features the vibrational and vibronic spectra, as an extension of the previous case study, for $n$=43, following the same structure and description as figure (\ref{fig4}). The $n$=43 case is of comparative value, as the system is not in the vicinity of a CI (figure (\ref{fig3}b)) as opposed to the $n$=48 case. The PECs features a much wider avoided crossing ($R\sim 2700$ a$_0$) as portrayed in figure (\ref{fig5}), and the differential couplings, while prominent, are not as peaked as in the $n$=48 case. The ground states of the BH spectra and the exact spectra are shifted by $\sim$11 MHz and $6.5$ MHz respectively, relative to the BO ground state. The positive energy shift due to the non-adiabatic coupling does not vary significantly between the $n$=43 and $n$=48 cases. But the deeper wells of the $n$=43 PECs ensure that the energy splitting is of less relative importance, than for the $n$=48 case. The first two excited eigenstates of the exact spectrum portray significant state mixing, and the non-adiabatic bound states are more localized than the BO states. The BH states near the continuum threshold show lower energy splitting than the $n$=48 case, which is associated to the fact that $n$=43 has a less prominent diagonal correction, due to it's less prominent non-adiabatic coupling. Much like the previous case, the vibronic interaction does introduce the radial localization of states for $n$=43. The fourth, fifth and eighth excited states of the BO spectrum are all delocalized over multiple wells, in contrast to the corresponding well-localized non-adiabatic states. As a final note in our comparison, a scattering resonance state with no BO or BH counterpart is observed at energy $\sim -1.278$ GHz, well above the continuum threshold of $-1.287$ GHz.

It is worth noting that although the non-adiabatic couplings are highly $n$-dependent, the behavior of the diabatic curves do not vary significantly over $n$. Hence the positive energy shifts ($\sim$MHz), state localization and existence of resonance states for the exact spectrum are all commonly observed for $30\leq n\leq60$. The pattern of $n$ dependent singularities (figure (\ref{fig3}b)) are also repeated regularly for $30\leq n\leq60$. But as the energies of the PECs scale as $n^{-3}$, the contribution of non-adiabatic interaction effects to the complete molecular spectra, becomes more important for higher $n$-values.

\section{Conclusion and Outlook}    \label{Sec5}

We have investigated non-adiabatic and vibronic interaction effects in the sodium ultralong-range molecule, using the vibronic coupling between states of trilobite and $d$-state character. The oscillatory nature of the trilobite and $d$-state PECs along with the specific quantum defect splitting of sodium, ensure the formation of avoided crossings for a range of $n$ values. We observe that the avoided crossing between the adiabatic PECs introduce significant non-adiabatic couplings that can behave like a singularity for specific $n$, which is associated with the existence of CIs in synthetic dimensions. Irrespective of the form of the derivative couplings, the exact spectra feature pronounced differences from the BO spectra, including MHz scale positive energy shifts in eigenvalues, localized probability densities and the existence of scattering resonance states near the avoided crossings, with no adiabatic counterpart. While the Born-Huang correction does partially explain the energy shift and the state localization, it cannot account for the exact spectra and the resonance states. Hence, our results confirm that a coupled-channel approach is necessary to study the vibronic spectra of sodium ULRM, irrespective of the corresponding principal quantum number $n$.

 A direct extension of our current work would be to include the $p$-wave scattering terms as well as fine and hyper-fine interaction. Inclusion of these terms, although representing a challenge, can help obtain results that are empirically comparable, and be further developed to study non-adiabatic effects in spin interacting systems. Non-adiabatic interactions in such systems might give insight into spin-changing, and $l$-changing collision dynamics.
 
 Furthermore, the present study of ULRM also opens up new research opportunities in non-adiabatic wave-packet dynamics. Notably, the introduction of an electric field breaks the spherical symmetry of the system and facilitates the existence of conical intersections in spatial dimensions. This provides ample opportunity to probe and observe non-adiabatic dynamics near CIs, on timescales of microseconds and distances of the order of micrometers, which are accessible in present-day cold-atom laboratories. Lastly, empirical exploration of ULRMs till date are focused on heavier molecules formed by K, Rb, Cs, and Sr. We justify the utility of lighter molecules based on Na, as they feature strong non-adiabatic effects and a rich vibronic structure, and promote interest in the experimental study of sodium Rydberg molecules.

\begin{acknowledgments}
We acknowledge support from the Deutsche Forschungsgemeinschaft (DFG) within the priority program ``Giant Interactions in Rydberg Systems" [DFG SPP 1929 GiRyd project SCHM 885/30-2]. R.S.~is grateful to Dan Bosworth for beneficial discussions.
\end{acknowledgments}

\bibliographystyle{apsrev4-2}
\bibliography{references}
\end{document}